\documentstyle[12pt,epsfig]{article}

\large
\newcommand{\be}{\begin{eqnarray}}
\newcommand{\ee}{\end{eqnarray}}
\newcommand\del{\partial}

\begin{document}
\setlength{\baselineskip}{21pt}
\pagestyle{empty}
\vfill
\eject
\begin{flushright}
SUNY-NTG-98/15
\end{flushright}

\vskip 2.0cm
\centerline{\Large \bf  Thouless Energy and Correlations 
of QCD Dirac Eigenvalues}
\vskip 1.5cm
\centerline{J.C. Osborn and J.J.M. Verbaarschot}
\vskip 0.2cm
\centerline{\it Department of Physics and Astronomy, SUNY, 
Stony Brook, New York 11794}
\vskip 2cm

\centerline{\bf Abstract}

Eigenvalues and eigenfunctions of the QCD Dirac operator are studied for
gauge field configurations given by a liquid of instantons. We find that
for energy differences $\delta E$ 
below an energy scale $E_c$ the eigenvalue 
correlations are given by Random Matrix
Theories with the chiral symmetries of the QCD partition function. 
For eigenvalues near zero
this energy scale shows a weak volume dependence that
is not consistent with $E_c \sim 1/L^2$ which might be expected from 
the pion Compton wavelength and from the behavior of the Thouless 
energy in mesoscopic systems. However, the numerical value of $E_c$ for
our largest volumes is in rough agreement with estimates from 
the pion Compton wavelength.
A scaling behaviour consistent with
$E_c\sim 1/L^2$ is found in the bulk of the spectrum. 
For $\delta E> E_c$ the number variance shows a linear dependence
with a slope which is larger than the nonzero multifractality index
of the wave functions.  Finally, the average spectral density and the 
scalar susceptibilities are discussed
in the context of quenched chiral perturbation theory. We argue
that a nonzero value of the disconnected scalar susceptibility requires
a linear dependence of the number variance on $\delta E$.
 
\vfill
\noindent

\eject
\pagestyle{plain}

\noindent
\section{Introduction}
\vskip 0.5cm

Random matrix theories have been applied extensively to disordered mesoscopic
systems (for recent reviews we refer to \cite{HDgang,Beenreview,Montambaux}). 
In particular, correlations of energy eigenvalues have
been studied in great detail. One important energy scale that enters
in these studies 
is the Thouless energy, $E_c$, which is essentially the inverse 
diffusion time 
of an electron through the sample \cite{Altshuler}. 
If the diffusion constant is given by $D$
and the linear dimension of the sample is equal to $L$ the Thouless 
energy is given by
\be
E_c = \frac{\hbar D}{L^2}.
\ee
What has been found is that eigenvalues
closer than $E_c$ are correlated according to the
invariant random matrix ensembles. Because such correlations are characteristic
for classically ergodic and chaotic quantum system this regime is also known
as the ergodic energy domain. A second energy scale is given by $\hbar/\tau_e$,
where $\tau_e$ is the elastic collision time. The energy regime with
$\delta E > \hbar/\tau_e$ is known as the ballistic regime whereas the domain
$E_c < \delta E <  \hbar/\tau_e$ is known under the names of diffusive regime
or Altshuler-Shklovskii regime. The different energy regimes have 
received a great deal of attention in recent literature 
\cite{Altland,Braun,Aronov,yan,Guhr,kravtsov-lerner,altland-gefen}
(for further references we refer to 
\cite{HDgang,Beenreview,Montambaux}). 
A semiclassical interpretation in terms of the
return probability has been given in \cite{Imry}. 

A convenient way to measure the pair correlations of eigenvalues  
is by means of the number variance $\Sigma_2(n)$. 
This statistic is defined as the variance of the number of eigenvalues in
an interval containing $n$ eigenvalues on average. 
In the ergodic regime
the number variance is given by the invariant random matrix ensembles
$\Sigma_2(n) \sim (2/\beta \pi^2) \log(n)$. 
The behavior of the number variance in the diffusive is less well-established.
One proposal is \cite{Altshuler} that $\Sigma_2(n) \sim n^{d/2}$ (where
$d$ is the number of Euclidean dimensions) if the coupling is below the critical
value for a localization transition.
However, for critical values of the coupling constant, for which the 
localization length scales with the size of the sample, a linear 
dependence on $n$ is predicted, i.e., $\Sigma_2(n)\sim \chi n$.
In this regime, the slope has been related to the multifractality index
of the wave functions \cite{Chalker-kravtsov,kravtsov}.  

In this letter we investigate to what extent such scenarios are realized 
in QCD. In QCD the order parameter of the chiral phase transition is
related to the spectral density, $\rho(\lambda)$, 
of the Dirac operator by means of the 
Banks-Casher formula \cite{BC}, $\Sigma = \lim \pi \rho(0)/V$
(the space-time volume is denoted by $V= L^4$). For broken
chiral symmetry, $\Sigma \ne 0$, this results in eigenvalues 
near zero virtuality that are spaced as $\sim 1/V$.  It is therefore
natural to introduce the microscopic limit $V\rightarrow \infty$ such that
$u=\lambda V \Sigma$ is kept constant.  The
microscopic spectral density is then defined by \cite{SV},
\be
\rho_S(u) = \lim_{V\rightarrow \infty} \frac 1{V\Sigma} \langle
\rho(\frac u{V\Sigma})\rangle,
\label{rhosu}
\ee
where the spectral density is defined by
\be
\rho(\lambda) = \sum_k \delta(\lambda -\lambda_k),
\ee
and the brackets denote ensemble averaging.
There is ample evidence from lattice QCD \cite{Tilo,Ma,Markum} 
and instanton liquid simulations \cite{Vinst} that $\rho_S(u)$ 
 and other correlators 
on the scale of individual level spacings \cite{Halasz} are given by 
chiral Random Matrix Theory, i.e.
random matrix theories with the chiral symmetries of the QCD partition
function. However, at scales beyond a few eigenvalue spacings
in both instanton simulations \cite{Vinst} and lattice
QCD simulations \cite{Tilo,Ma} the Dirac eigenvalues 
near zero show
stronger fluctuations than those in the chiral random matrix theories.
This indicates the presence of an energy scale in QCD which may be 
identified as the Thouless energy.

In the regime where the pion loops can be ignored the 
correlations of the eigenvalues near zero are given by the
chiral random matrix ensembles \cite{SV,V}. They can be mapped onto   
an effective partition function given by the mass term generated by the
spontaneous breaking of chiral symmetry \cite{SV}. 
From the work of Gasser, Leutwyler and Smilga \cite{GL,LS}, 
we expect that the boundary of the ergodic regime is given by
a mass scale $m_c$ where the Compton wavelength of the associated Goldstone 
boson with mass $m_{\pi}^2 = m_c K$ (according to the PCAC relation
$K = \Sigma/F_\pi^2$, where $F_\pi$ is the pion decay constant)
is of the order of the linear dimension
of the box. This condition can be written as \cite{vPLB,Trento}
\be
m_c = \frac 1{K L^2}.
\label{range}
\ee
It is therefore tempting to interpret $1/K$ as the diffusion constant.
Because $K$ is large on a hadronic scale ($K\approx 1660\, MeV$) the diffusion
constant is relatively small, and we expect a Thouless energy that is small
as well.  We emphasize that $m_c$ is a valence quark mass that is not
present in the fermion determinant. In other words, the spectral density
does not depend on $m_c$. For example, the valence quark mass of the
chiral condensate is related to the spectral density by \cite{Christ}
\be
\Sigma(m) = \frac 1V
\int_0^\infty \frac {2m\rho(\lambda) d\lambda }{\lambda^2 +m^2}.
\label{valence}
\ee
Therefore, the spectral density and, similarly, higher order correlation 
functions, can be obtained from the valence quark mass dependence 
of the effective partition function. 
It can be obtained with the help of quenched chiral perturbation theory
\cite{Golterman}
or by means of the replica trick in which we start from $n_f$ valence flavors
and take the limit $n_f\rightarrow 0$ at the end of the calculation. 
In the thermodynamic limit, the mass scale (\ref{range}) is much less than the
scale of  massive quarks which leads to a decoupling of the valence quark mass
dependent part of the partition function. 
The Dirac spectrum and its correlations are 
thus given by the valence quark mass
dependence of the effective partition function which below the scale 
(\ref{range}) is given by chiral random matrix theory. 
For massless quarks the argument is simpler. Now the 
Compton wavelength of the physical pions is always much larger than the 
size of the box. The crossover point is thus given by the value of the
valence quark mass where the corresponding Goldstone boson 
(in the sense of quenched or partially quenched chiral perturbation
theory) is equal to the size of the box. In this case, however, the 
microscopic spectral correlations depend on the number of massless flavors.
Alternatively, one can study the boundaries of the domain where the pion loops
can be neglected via  relations between microscopic  spectral correlators
and finite volume partition functions as given in \cite{Dampart}.

In the bulk of the spectrum, chiral symmetry is broken explicitly by
the valence quark mass, and the 'Goldstone bosons' associated with
the valence quark masses  are no longer in the low energy spectrum of
QCD. However, in this case, spontaneous symmetry breaking in the 
generating function for the correlation functions results in the usual
nonlinear $\sigma-$ model for  disordered mesoscopic systems \cite{Efetov}.
The critical energy scale is then given by the Thouless energy
$m_c \sim 1/K' L^2$, where $K' = \pi\rho(E)/VF_\pi^2$ and $\rho(E)$ is the
spectral density at $E$. Since the group manifold is different in this case 
\cite{Andreev}
the numerical constant in this relation differs by a factor of the order unity
from the constant in (\ref{range}).

Using that according to the Banks Casher relation \cite{BC} the
eigenvalue spacing is given by $\Delta = \pi/\Sigma V$, the condition 
(\ref{range}) can
be rewritten in dimensionless form as
\be
g_c = \frac{m_c}\Delta = \frac{F_\pi^2}\pi L^2.
\label{gc}
\ee
Here, $g_c$ plays the role of the dimensionless conductivity. At this point
we wish to mention that a relation between transport properties 
of mesoscopic systems and the pion decay constant was established recently
in \cite{Stern}.
In lattice QCD
for an $Na^4$ lattice this relation reads $g_c = F_\pi^2 a^2 \sqrt N/\pi$. On a
$16^4$ lattice with a lattice spacing of 0.1\,$fm$ this results in a
dimensionless Thouless energy of a couple of lattice spacings.  

In order to investigate the presence of a Thouless energy in QCD
we perform our calculations for gauge field configurations 
given by a liquid of instantons. For such configurations we
study correlations of the eigenvalues of the corresponding Dirac operator. 
In particular, we investigate the number variance, and establish a crossover
between the  ergodic
and the diffusive domain. The linear dependence will be compared with the 
multifractality index of the wave functions. We will argue that results
for chiral perturbation theory apply to the diffusive regime.
The above observables are studied for a variety of system sizes and as
a function of the number of flavors.

\section{Instanton liquid model}

In this model the 
gauge field configurations are given by a superposition of instantons.
We use the streamline Ansatz which for a dilute system
reduces to the sum Ansatz defined as a simple superposition of 
instanton profiles $A_\mu = \sum_I A_{I,\mu}$. Each instanton is described by
$4N_c$ (the number of colors is denoted by $N_c$) collective coordinates. The
Euclidean QCD partition function is then approximated by
\be
Z_{\rm inst} = \int D\Omega \, {\det}^{N_f} (\gamma\cdot D + m) e^{-S_{\rm YM}},
\label{zinst}
\ee
where the integral is over the collective coordinates of the instantons
and the covariant derivative is defined by $D_\mu=\del_\mu+iA_\mu $. 
The Yang-Mills action of the gauge field configurations is denoted by 
$S_{\rm YM}$. We will
only consider configurations with an equal number of instantons and 
anti-instantons. The fermion determinant is evaluated in the space
of the fermionic zero modes of the instantons.

This partition function obeys the flavor and chiral symmetries of the QCD
partition function. It reproduces results obtained 
in chiral perturbation theory as well as many other details of the
low energy hadronic phenomenology \cite{mitjarus,mitjanpb,mitjapre,SVinst}. 
In our calculations we use the standard instanton liquid
parameters with instanton density $N/V = 1$ (in units of $fm^{-1}$).
For more details on the instanton liquid
partition function we refer to a recent review by Sch\"afer and Shuryak
\cite{SS97}. 

\section{Random matrix model}
The chiral random matrix ensembles for $N_f$ massless quarks in the sector
of topological charge $\nu$ are defined by
the partition function \cite{SV,V}
\be
Z_{N_f,\nu}^\beta =
\int DW  {\det}^{N_f}\left (\begin{array}{cc} 0 & iW\\
iW^\dagger & 0 \end{array} \right )
e^{-{n \beta} {\rm Tr}V(W^\dagger W)},
\label{zrandom}
\ee
where $W$ is a $n\times (n+\nu)$ matrix.
As is the case in QCD, we assume that
$\nu$ does not exceed $\sqrt {2n}$. The parameter $2n$
is identified as the dimensionless volume of space time. It can be thought
of as the total number of instantons in the
partition function (\ref{zinst}).
The matrix elements of $W$ are either real ($\beta = 1$, chiral
Orthogonal Ensemble (chOE)), complex
($\beta = 2$, chiral Unitary Ensemble (chUE)),
or quaternion real ($\beta = 4$, chiral Symplectic Ensemble (chSE)).
The simplest ensemble is the Gaussian case with $V(x) = \Sigma^2 x$
(also known as the Laguerre ensemble).
In that case one can easily show that the microscopic spectral density
(\ref{rhosu}) is given by \cite{V}
\be
\rho_S(u)  = \frac {u}{2} (J^2_{a}(u) -J_{a+1}(u)
             J_{a-1}(u)).
\label{micro2}
\ee 
where $a= N_f + |\nu|$ (the result for $a=0$ was obtained in \cite{VZ}).
It is also straightforward to derive the microscopic correlation functions
\cite{VZ,Ma}. In the bulk of the spectrum, 
the eigenvalue correlations of the chiral ensembles
are given by the invariant random matrix ensembles.

It was shown by Akemann et al. \cite{Damgaard} that, for $\beta =2$,
the microscopic spectral density and the
microscopic spectral correlators do not depend on the potential $V(x)$
and are given by the results \cite{VZ} for the Laguerre ensemble. Their 
work greatly extends an earlier result by Br\'ezin and Zee \cite{Brezin-Zee}
who studied
the potential $V(x) = x + \alpha x^2$.
The microscopic
spectral density is also invariant with respect to the addition of an 
arbitrary constant matrix to the off-diagonal blocks in the determinant
\cite{Tilo-Guhr,Seneru}. The application of chRMT has been 
put on a firm foundation
by these and other universality proofs \cite{other}. Whether or not QCD is
in this universality class is a dynamical question that only can be proven
by explicit numerical simulations.

\section{Numerical Results}

In this section we simulate the instanton liquid model introduced in 
section 2. We evaluate the partition function (\ref{zinst}) by means of
a Metropolis algorithm. Typically, we perform on the order of 10,000
sweeps for each set of parameters. For the equilibrated configurations
the eigenvalues and eigenvectors of the Dirac operator are calculated by
means of standard diagonalization procedures. 
Most of our calculations have been performed in the quenched approximation
($N_f=0$). The motivation for this choice is twofold. First, calculations
at $N_f \ne 0$ require the evaluation of a determinant for each update of
the integration variables which is extremely costly and restricts us
to a total number of instantons not larger than 128. For $N_f=0$, 
we are able to generate ensembles up to 512 instantons which turns out
to be important in the study of the thermodynamic limit.
Second, our results are 
compared with ideas from disordered mesoscopic systems where no  
fermion determinant is present. 

In order to separate the fluctuations of the eigenvalues from
the average spectral density we unfold the spectrum, i.e. we rescale the
spectrum in units of the local average level spacing. Specifically, the
unfolded spectrum $\lambda_k^{\rm unf}$ is 
obtained from the original spectrum $\lambda_k$ according to 
\be
\lambda_k^{\rm unf} = \int_0^{\lambda_k} \bar\rho(s) ds,
\ee 
where $\bar\rho(s)$ is the smoothened average spectral density. 
The number variance,
the nearest neighbor spacing distribution and the microscopic spectral
density are all calculated from the unfolded spectrum. 

\begin{center}
\begin{figure}[!ht]
\centering
\includegraphics[width=75mm]{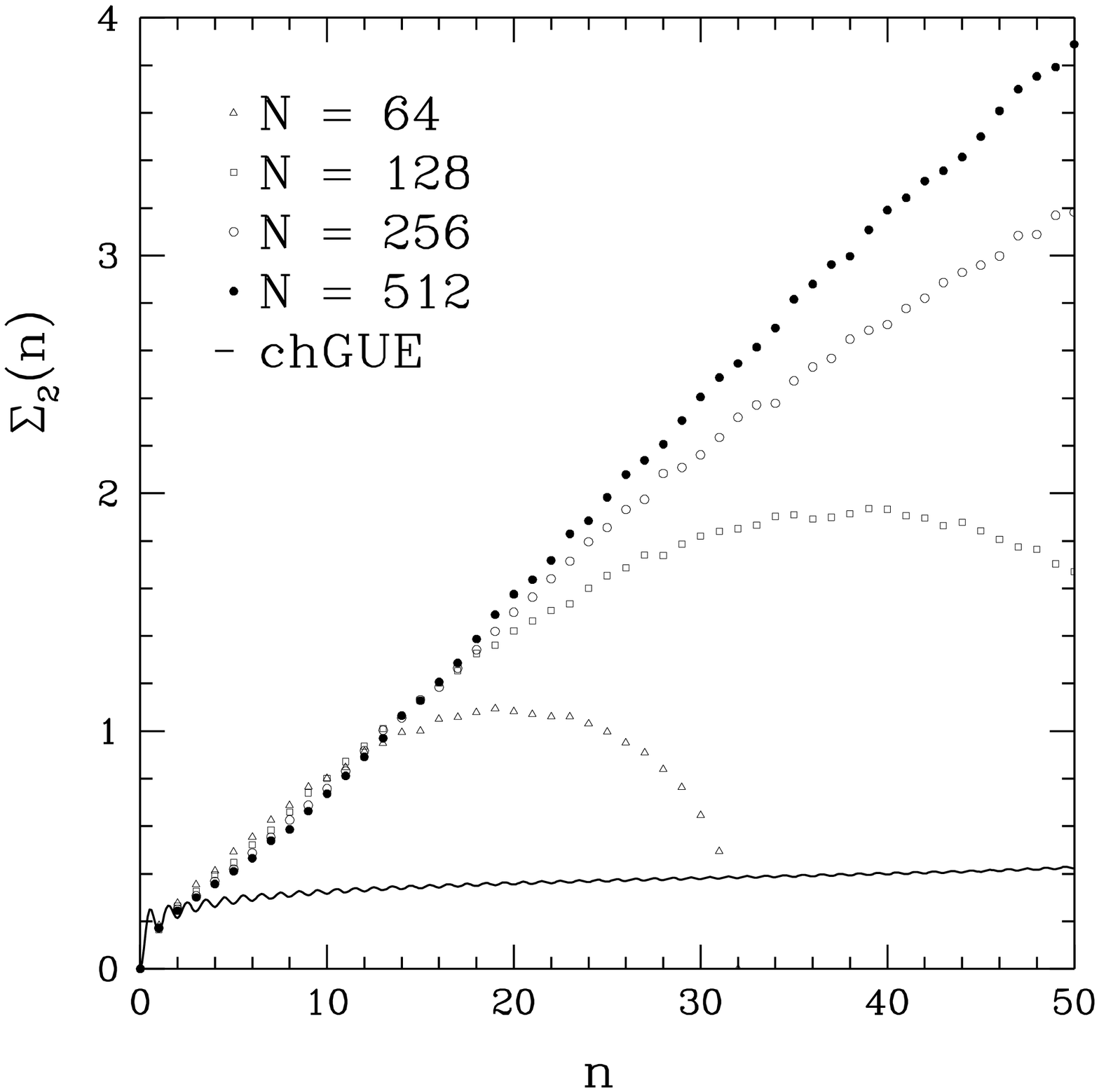}
\includegraphics[width=75mm]{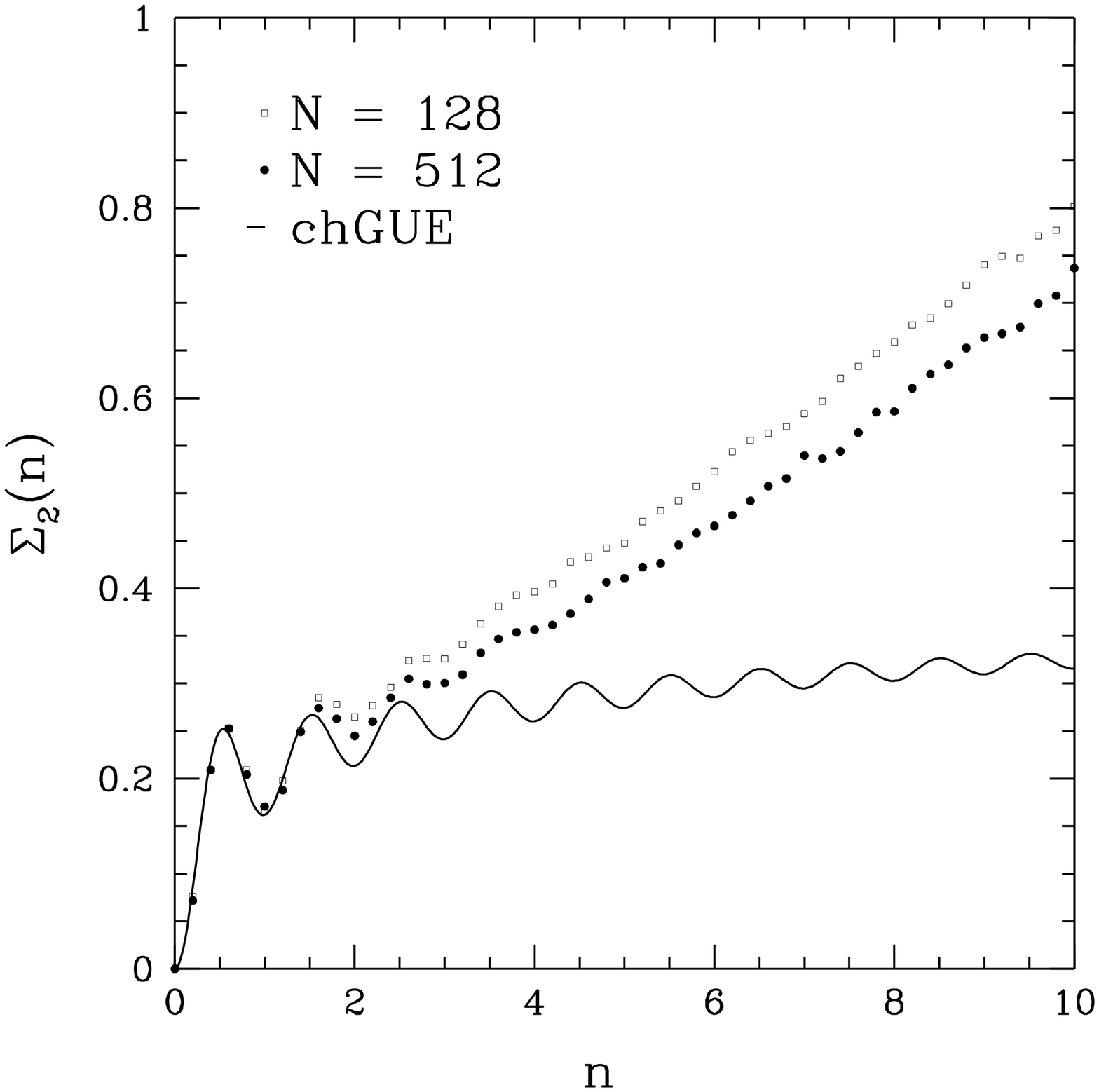}
\vspace*{0.3cm}
\begin{center}
\begin{minipage}{13cm}
\baselineskip=12pt
{\begin{small}
Fig. 1. The number variance $\Sigma_2(n)$ versus $n$ in the quenched
approximation for an interval starting at $\lambda = 0$. The
total number of instantons is indicated in the
legend of the figure.
\end{small}}
\end{minipage}
\end{center}
\vspace*{-0.1cm}
\end{figure}
\end{center}

In Figs. 1 and 2 we show the number variance, $\Sigma_2(n)$ versus $n$ 
for $N_f = 0$ and various total numbers of 
instantons as given in the legend of the figure. The 
instanton density $N/V= 1$. 
The random matrix result for $\Sigma_2(n)$  for the chGUE \cite{Ma} is
depicted by the solid curve. In both figures, the
right figure (with results for only two different volumes) 
is a blown-up version of the left figure.
In Fig. 1 the number variance is calculated for the interval that contains
on average the $n$ smallest positive eigenvalues. Fig. 2 represents the 
number variance in the bulk of the spectrum obtained from an interval that
is symmetric about the average unfolded positive eigenvalue. In both cases
we observe a transition point $n_c$ below which the number variance
is given by random matrix theory.
In Fig. 1  the value of the crossover point, $n_c\approx 2$, 
depends only weakly on  the total number of instantons
(or the volume). This is not in agreement with
the theoretical expectation (\ref{gc}) 
that $n_c \approx F_\pi^2 \sqrt V/\pi$ in four dimensions. 
For standard instanton
liquid parameters the numerical value of $n_c$ for $N$ instantons is given by
$n_c \approx 0.08 \sqrt N$ which is on the order of the results found in Fig. 
1. In the bulk of the spectrum (Fig. 2) the value of $n_c$ is
consistent with a $\sqrt V$ scaling but the numerical constant appears 
to be larger than the above estimate (even if we take into account that
the spectral density in the bulk is considerably less than near $\lambda =0$).
This result is in agreement with the finding that 
correlations of lattice QCD eigenvalues for  $12^4$ and $16^4$ lattices
are given by RMT up to distances of 100 and 300 level spacings, 
respectively \cite{Halasz,Guhrp}. Notice that the condition that the 
pion Compton wavelength is equal to the size of the box (leading to 
(\ref{range})) provides us only with
an order of magnitude estimate for the transition point.

\begin{center}
\begin{figure}[!ht]
\centering
\includegraphics[width=75mm]{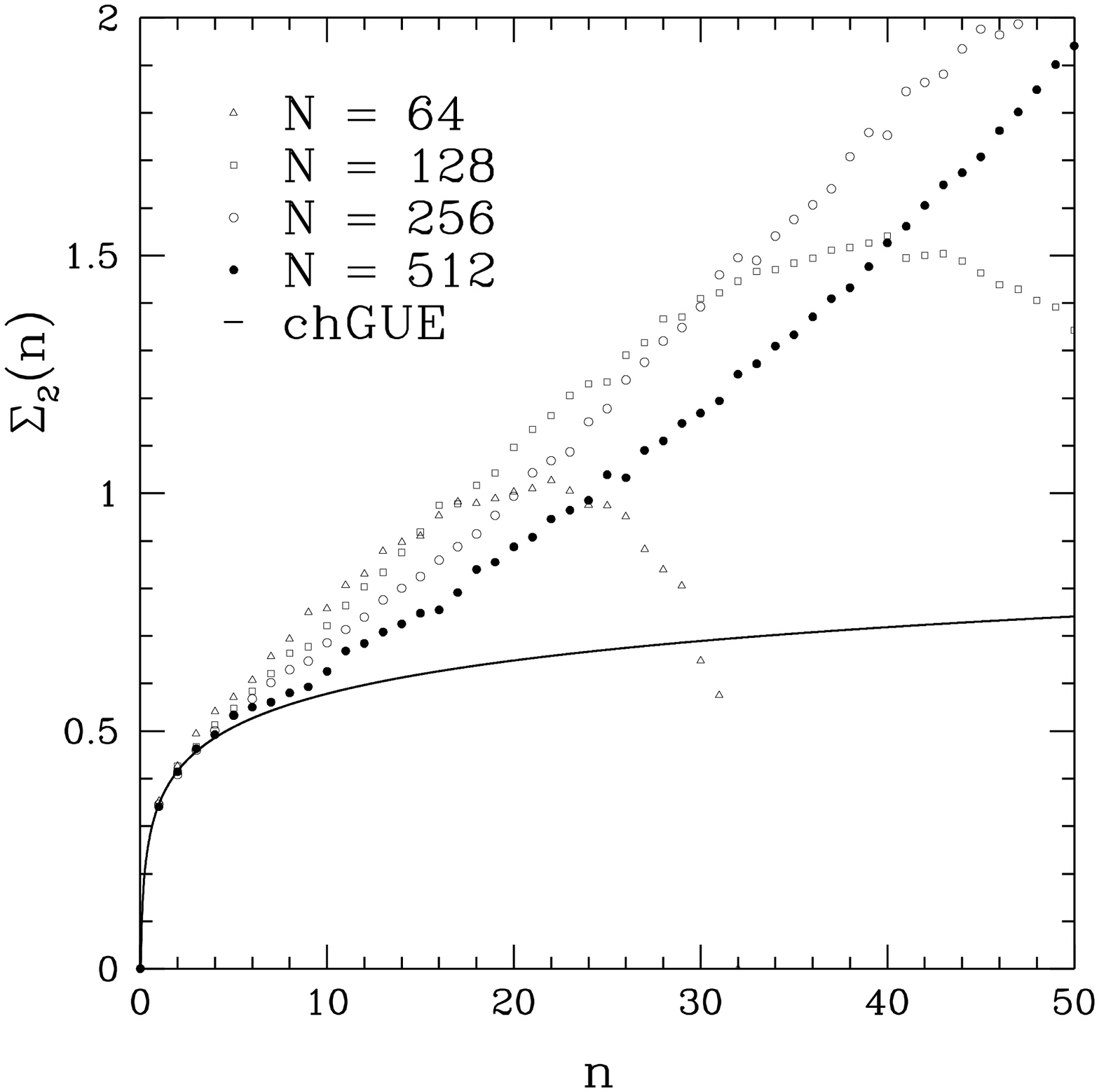}
\includegraphics[width=75mm]{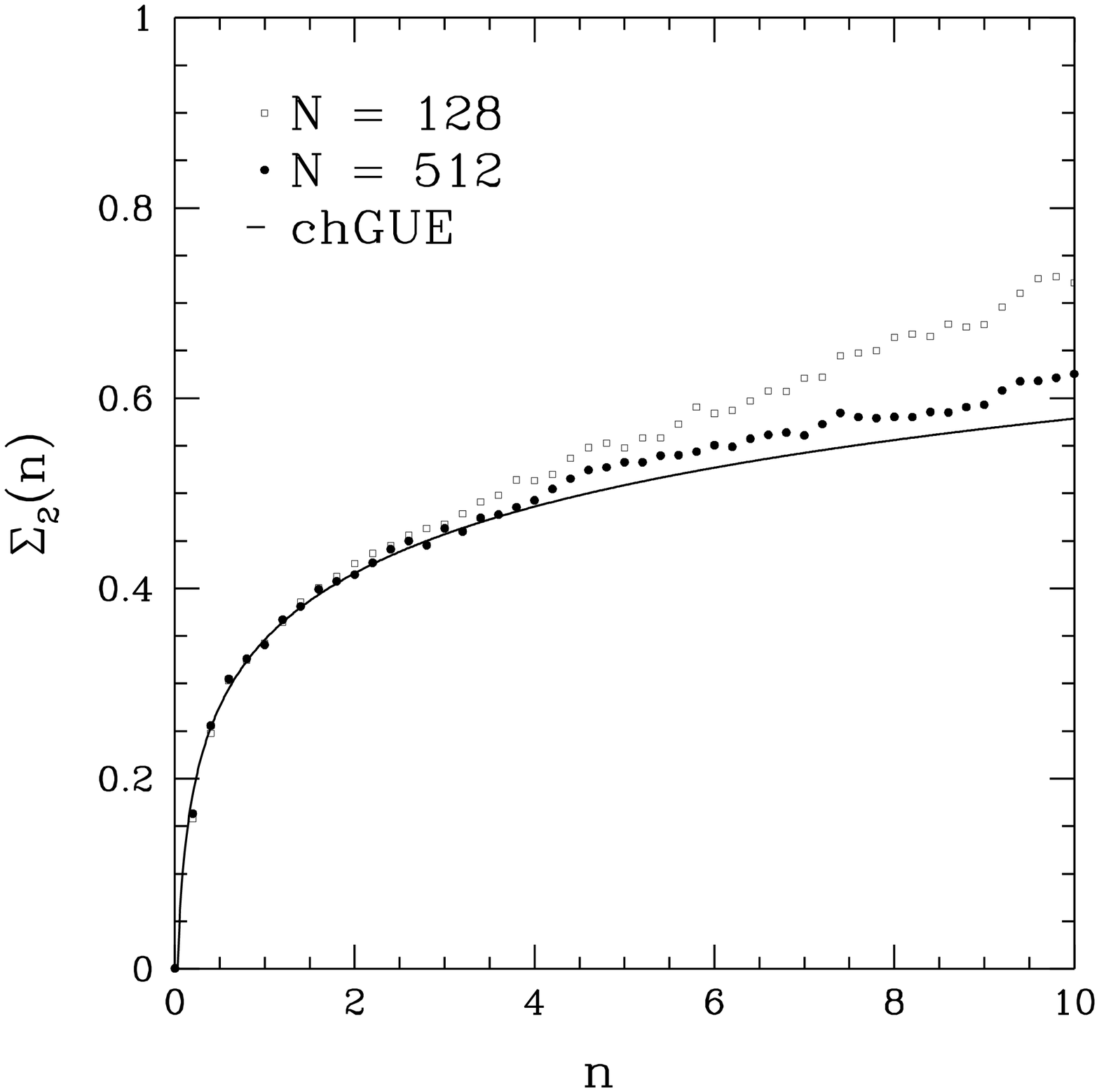}
\vspace*{0.3cm}
\begin{center}
\begin{minipage}{13cm}
\baselineskip=12pt
{\begin{small}
Fig. 2. The number variance $\Sigma_2(n)$ versus $n$ in the bulk of the
spectrum in the quenched
approximation and a total number of instantons as indicated in the
legend of the figure.
\end{small}}
\end{minipage}
\end{center}
\vspace*{-0.1cm}
\end{figure}
\end{center}

Beyond the crossover point the number variance shows 
a linear behavior with a slope 
$\chi \approx 0.08$ for eigenvalues near zero and $\chi \approx 0.04$ 
in the bulk of the spectrum. 
The downward trend of the curves for larger values of $n$
is a finite size effect. 
For example, for a finite number of uncorrelated eigenvalues $\Sigma_2(n)$
does not behave linearly, but rather as $\Sigma_2(n) = n-2n^2/N$
(notice that we have only $N/2$ positive eigenvalues).
This finite size effect prevents us from  saying
more about the ballistic regime, an energy scale of roughly 
the inverse distance between instantons. 

\begin{center}
\begin{figure}[!ht]
\centering
\includegraphics[width=70mm]{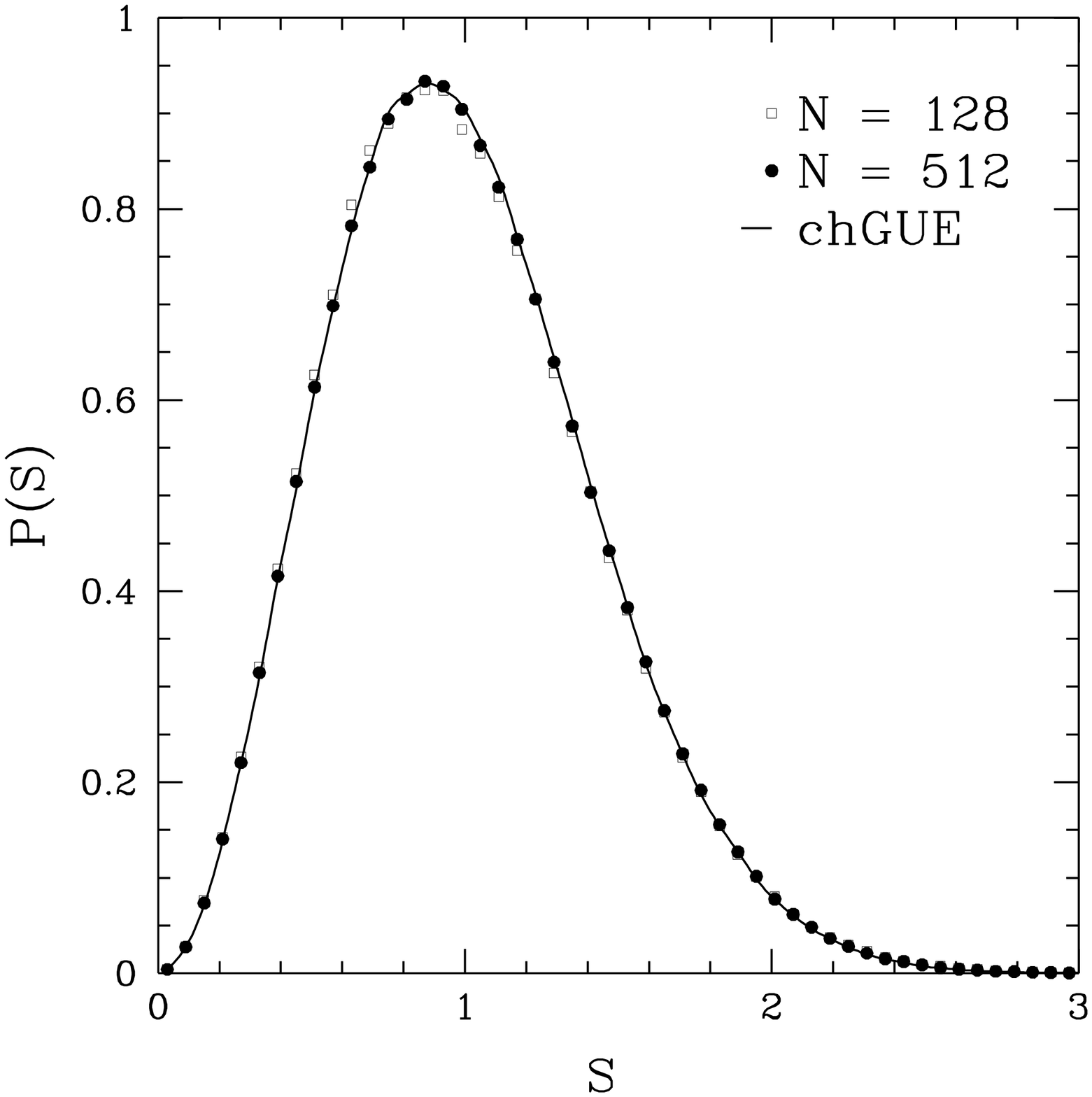}
\includegraphics[width=70mm]{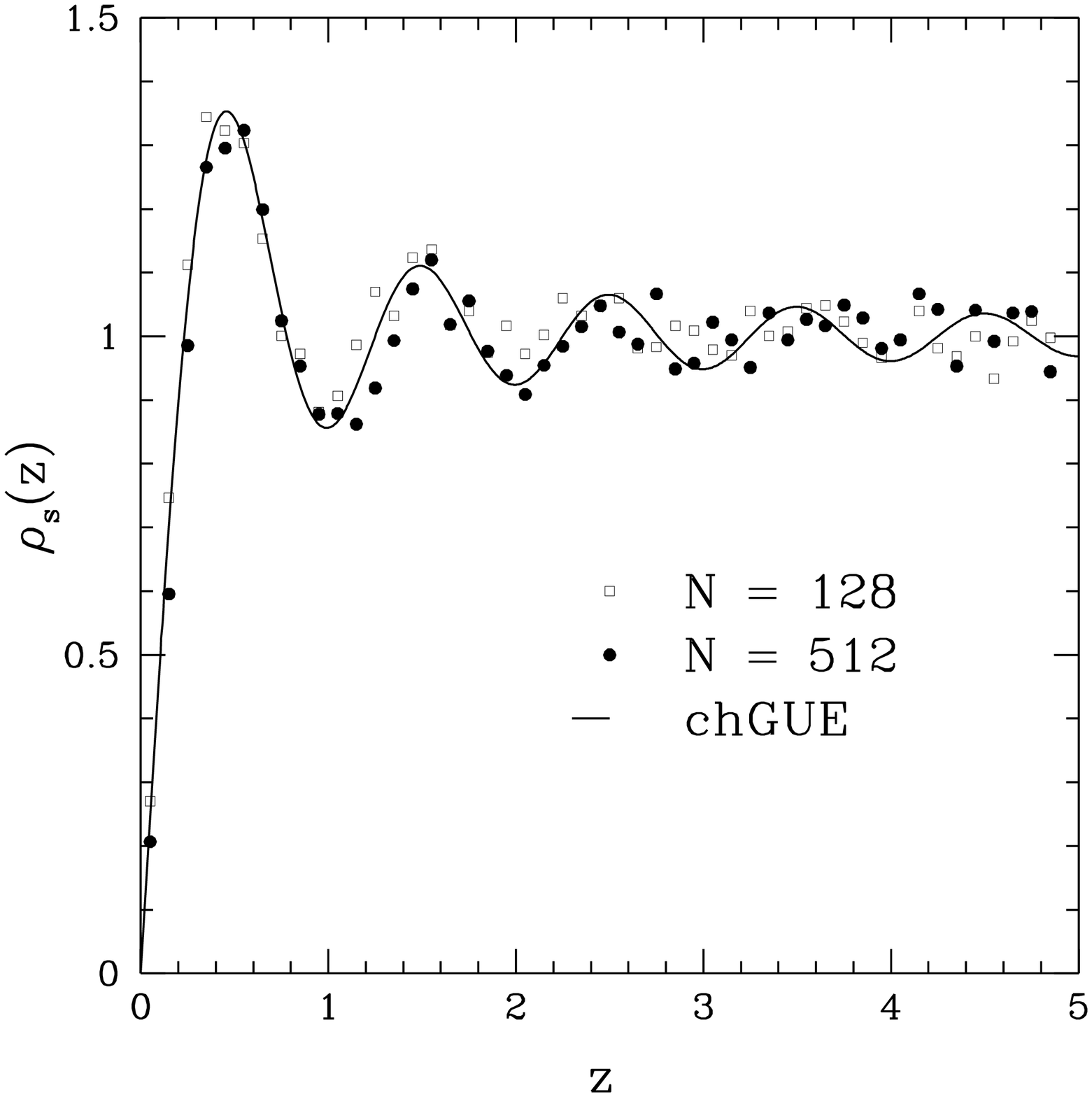}
\vspace*{0.3cm}
\begin{center}
\begin{minipage}{13cm}
\baselineskip=12pt
{\begin{small}
Fig. 3. The nearest neighbor spacing distribution 
$P(S)$ versus $S$ (left) and the
microscopic spectral density $\rho_S(z)$ versus $z$ (right). 
The full circles and the open squares represent results
for 512 and 128 instantons, respectively. Results obtained from the 
(chiral) random matrix ensembles are depicted by the full line.
\end{small}}
\end{minipage}
\end{center}
\vspace*{-0.1cm}
\end{figure}
\end{center}

In the ergodic regime we expect that eigenvalue correlations are
given by the invariant random matrix ensembles. This is shown in Fig. 3 where
we plot the nearest neighbor spacing distribution 
$P(S)$ versus $S$ (left) and the
microscopic spectral density $\rho_S(z)$ versus $z$ (right) (in terms of
the variable $z = u/\pi$ in (\ref{rhosu})).
The full circles and the open squares represent results
for 512 and 128 instantons, respectively. Results obtained for the 
(chiral) random matrix ensembles are depicted by the full line.
Clearly, below the Thouless energy we are in close agreement with RMT.
However, beyond the Thouless energy of about two level spacings
the oscillatory structure in the microscopic spectral density
is no longer reproduced. This implies that beyond this point
the QCD Dirac eigenvalues fluctuate more than the chGUE eigenvalues which
is  consistent with the number variance shown in Fig. 1. 
A similar increase in eigenvalue fluctuations at a scale of a few
eigenvalue spacings has been observed in
lattice QCD spectra \cite{Tilo} indicating that also in that case the Thouless
energy for the region around $\lambda = 0$ 
is less than predicted by the scaling behavior of disordered 
mesoscopic systems.

The inverse participation ratio which is a measure for  the number of
significant components in the wave function 
is defined as 
\be
I_2(\lambda) = \langle \sum_k |\psi_k(\lambda)|^4 \rangle,
\ee 
The $\psi_k(\lambda)$ are the normalized eigenfunctions of the Dirac operator
in the space of fermionic zero modes corresponding to 
eigenvalue $\lambda$. The number of components of the wave function will be
denoted by $N$.
For the chGUE we find $I_2 = 2/( N+2)$ whereas for uncorrelated eigenvalues
(the chiral Poisson ensemble) we find $I_2 = 1/2$.
Results for $N I_2(\lambda)$ as a function of 
$\lambda$ are depicted in Fig. 4 (left). 
The general impression
is that the wave-functions are extended with a participation ratio that is
not too different from the random matrix result (full line).
The eigenfunctions corresponding to small and large eigenvalues appear
to be somewhat more localized. A more definitive result for 
the character of the
wavefunctions follows from the scaling behavior of 
$I_2(\lambda)$ with the volume. 
\begin{center}
\begin{figure}[!ht]
\centering
\includegraphics[width=75mm]{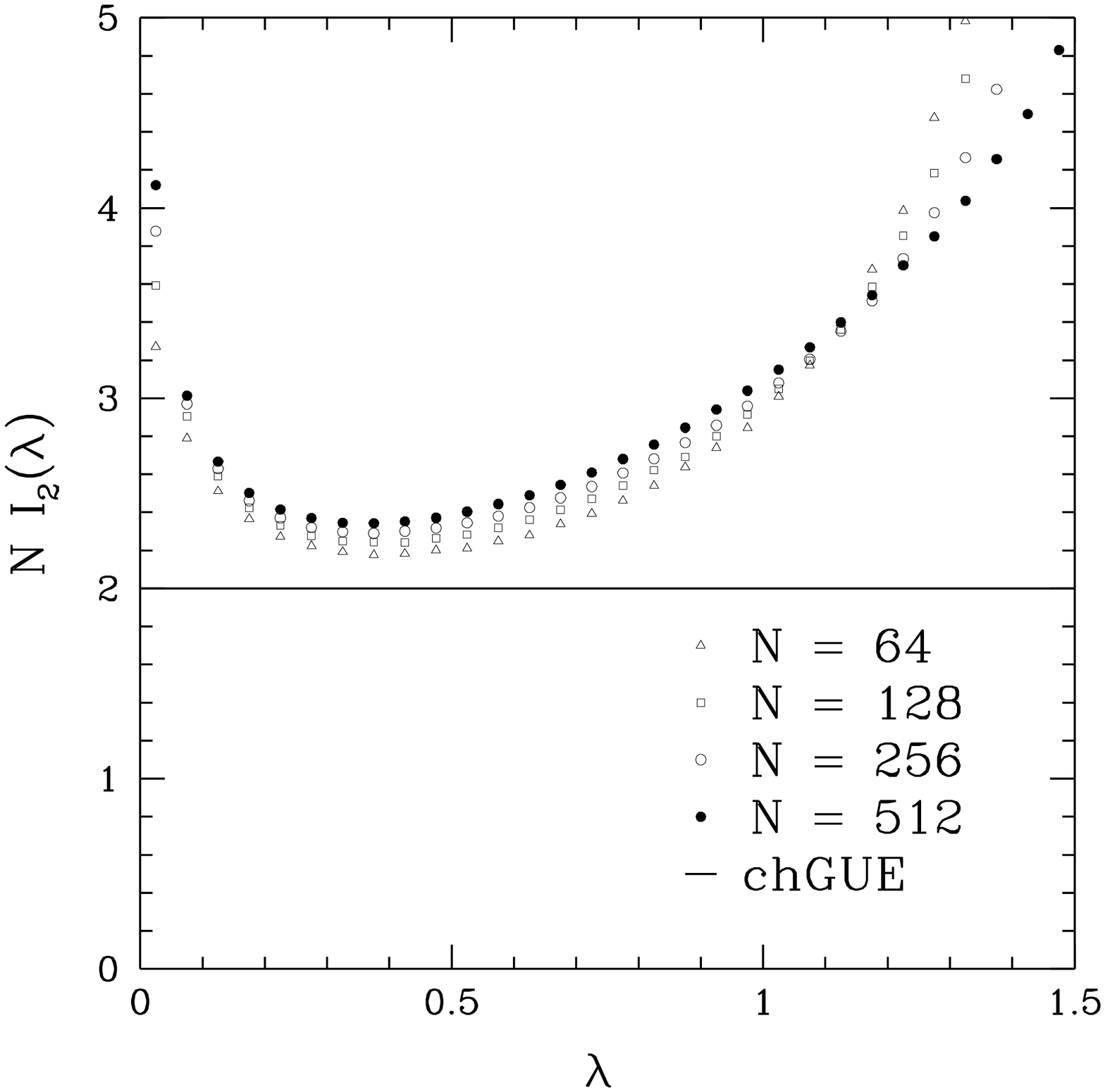}
\includegraphics[width=75mm]{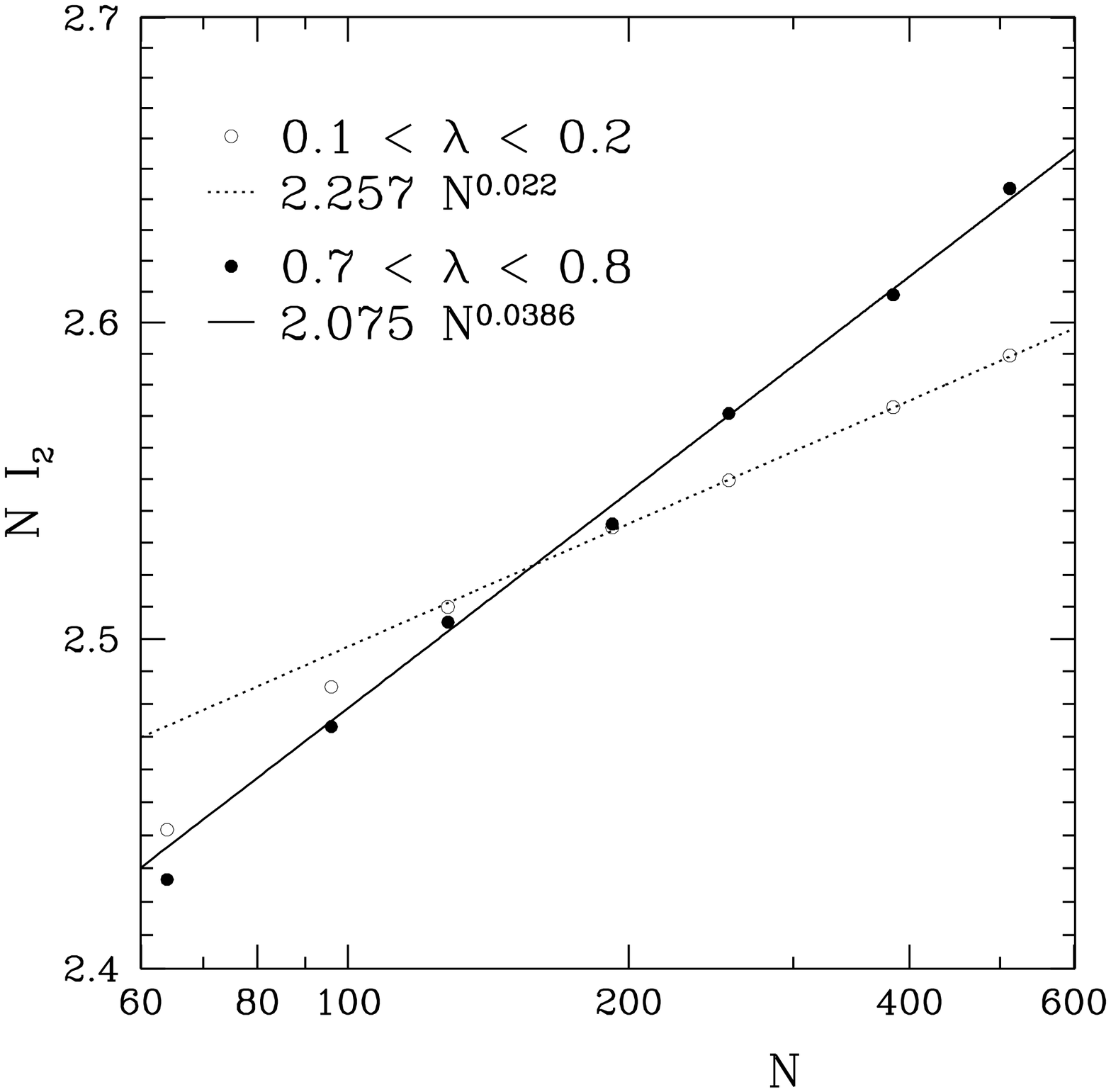}
\vspace*{0.3cm}
\begin{center}
\begin{minipage}{13cm}
\baselineskip=12pt
{\begin{small}
Fig. 4. The inverse participation ratio times $N$ 
versus the corresponding Dirac 
eigenvalues for $N_f =0$ (left) and a total number of instantons equal to
64, 128, 256 and 512, and the scaling behavior versus $N$ (right) 
for the energy intervals $[0.1, 0.2]$ and $[0.7,0.8]$.
\end{small}}
\end{minipage}
\end{center}
\vspace*{-0.1cm}
\end{figure}
\end{center}
A double logarithmic plot of 
$NI_2(\lambda)$ versus $N$ is  shown in Fig. 4 (right). Results are given
both for the energy intervals $[0.1,0.2]$ (open circles) and $[0.7,0.8]$ (full 
circles). The first region corresponds to a part of the spectrum where
the number variance shows a linear behavior for the volumes shown in Fig. 1,
and the second region corresponds to the bulk of the spectrum.

The multifractality
index $\eta$ of the wave functions is defined by \cite{Chalker-kravtsov} 
\be
I_2 \sim V^{\eta/d-1}.
\ee
According to an argument given in \cite{Chalker-kravtsov} the 
value of $\eta=2\chi d$ (where $\chi$ is the slope of the 
linear piece in the number variance) in the critical domain. 
From the volume dependence of the inverse participation ratio shown in 
Fig. 4 (right)
we find a value
for $\eta/d$ of about 0.02 and 0.04 for the intervals
$[0.1,0.2]$ and $[0.7,0.8]$, respectively. These values are well below
the theoretical result of $2\chi$. Apparently, the ensemble of instantons
is not in the critical region for $N_f =0$. The localization properties
of eigenfunctions have also been studied for the Wilson lattice QCD
Dirac operator \cite{Janssen}. 
In that case it was found that the eigenstates are localized.
We have no explanation for this discrepancy.

The flavor dependence of the number variance of eigenvalues near zero
is shown in Fig. 5. 
Results are given for 128 instantons and massless flavors.
Again the Thouless energy is about two level spacings (in order to avoid
a cluttered figure we have not plotted the random matrix results for
the number variance).
We observe an enhancement of eigenvalue fluctuations for 3 flavors
(chiral symmetry is already restored for $N_f = 4$ \cite{SS97}) and conclude 
that the critical number of flavors is approximately equal to three. 
However, our volumes (of up to 128 instantons) 
were too small to reliably extract the multifractality index.
For completeness we mention that a similar enhancement of fluctuations 
for  $N_f = 3$ also takes place in the bulk of the spectrum.

\begin{center}
\begin{figure}[!ht]
\centering
\includegraphics[width=100mm]{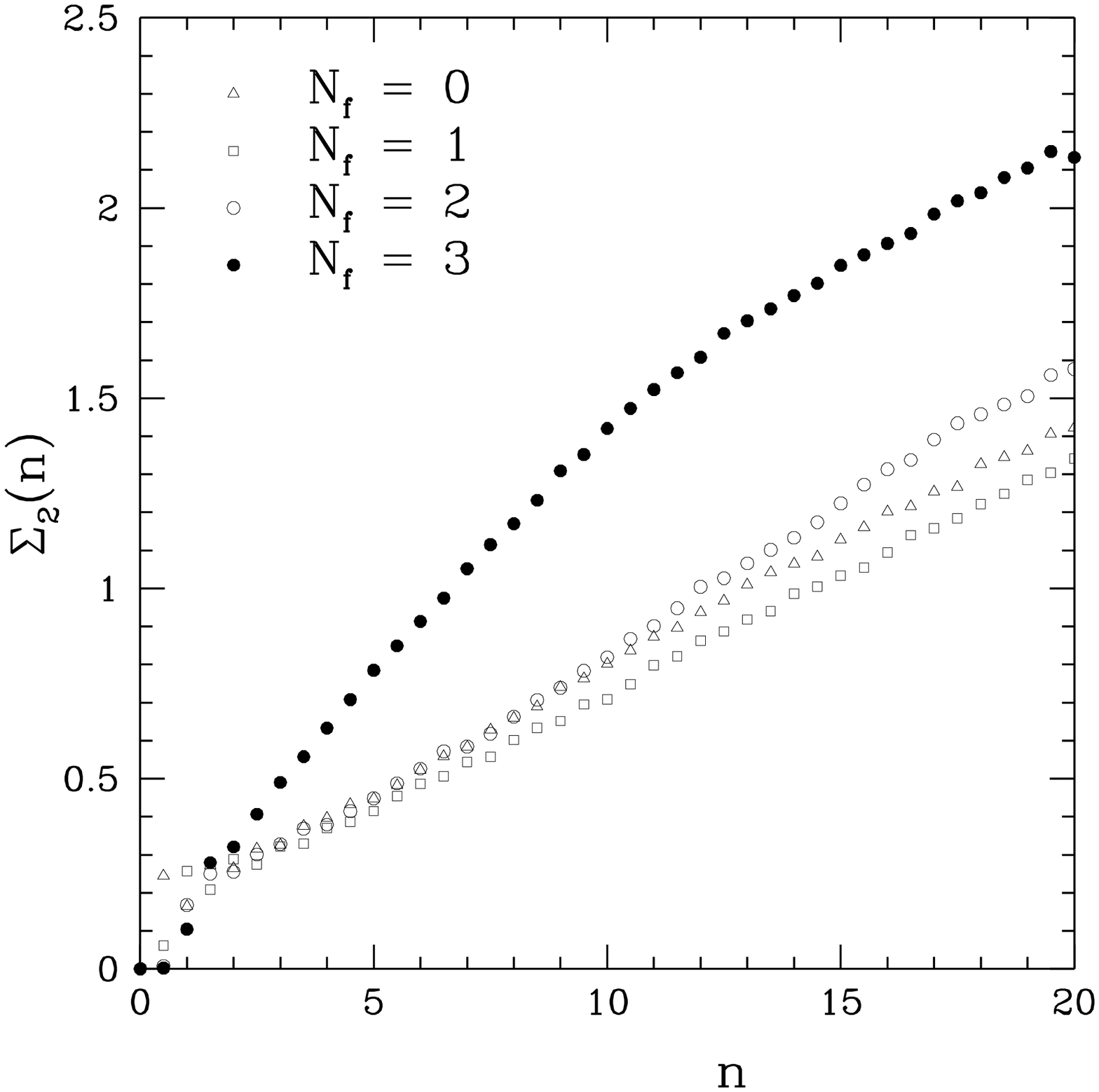}
\vspace*{0.3cm}
\begin{center}
\begin{minipage}{13cm}
\baselineskip=12pt
{\begin{small}
Fig. 5. The flavor dependence of the number variance $\Sigma_2(n)$
versus $n$ calculated for an ensemble of 128 instantons.
\end{small}}
\end{minipage}
\end{center}
\vspace*{-1cm}
\end{figure}
\end{center}

\section{Spectral density and chiral susceptibility}

In this section we consider the average spectral density, the valence
quark mass dependence of the chiral condensate and the 
connected and disconnected scalar susceptibilities. 
In order to compare our results with quenched chiral perturbation 
theory \cite{Golterman} we put particular emphasis 
on the extrapolation to the thermodynamic limit. 
These quantities were calculated elsewhere
for smaller volumes (64 instantons) \cite{Trento,Thomas}. 

\begin{center}
\begin{figure}[!ht]
\centering
\includegraphics[width=100mm]{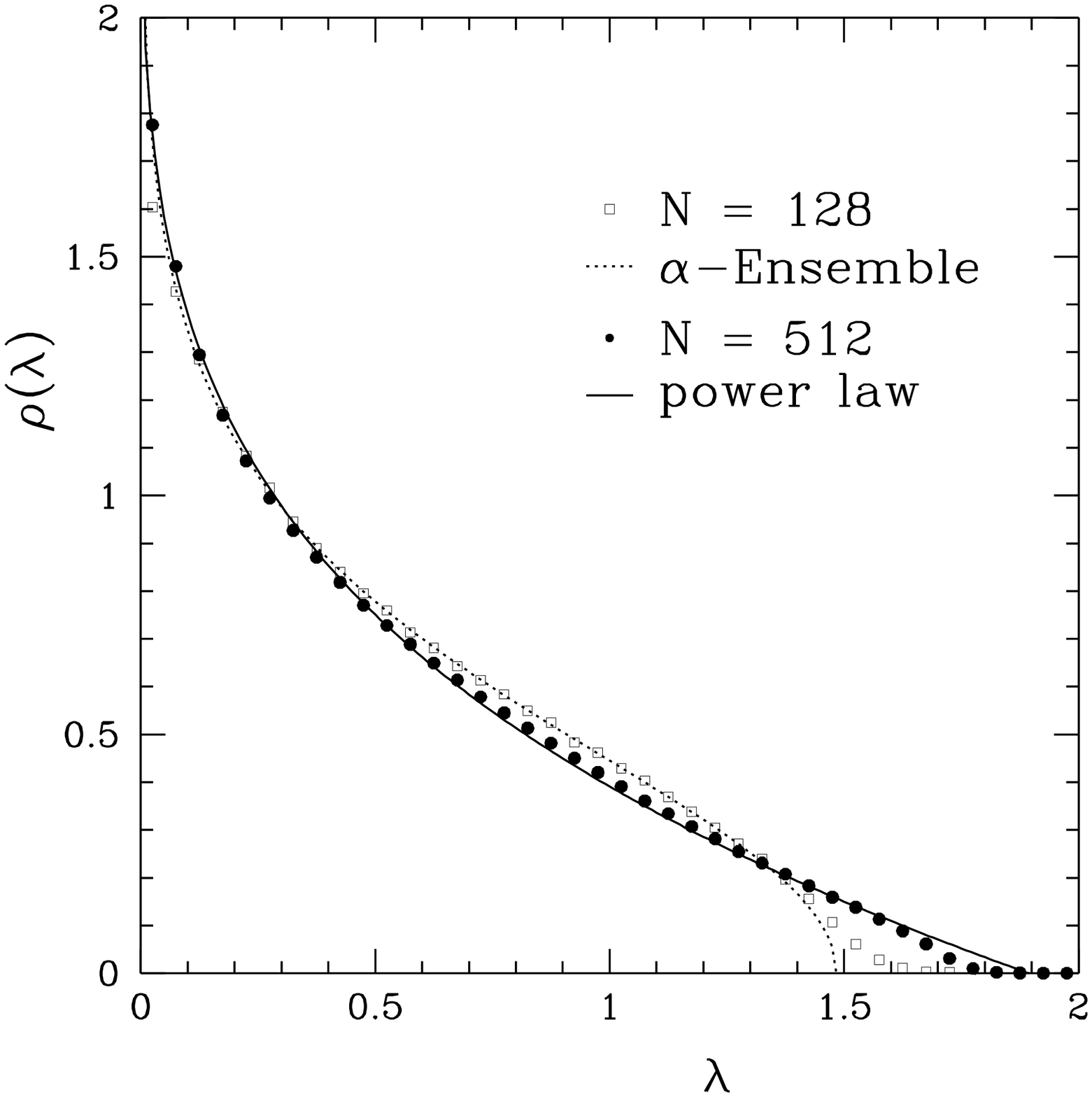}
\vspace*{0.3cm}
\begin{center}
\begin{minipage}{13cm}
\baselineskip=12pt
{\begin{small}
Fig. 6. The normalized spectral density 
$\rho(\lambda)$ versus $\lambda$ for $N_f =0$
and 128 and 512 instantons.
\end{small}}
\end{minipage}
\end{center}
\vspace*{-0.1cm}
\end{figure}
\end{center}

The spectral density of the Dirac eigenvalues for  a liquid of instantons was
studied before for $N_f = 0$ with the help of  mean field techniques 
\cite{mitjanpb,mitjarus}. For
large values of $N_c$ the overlap matrix elements are independent resulting
in a semi-circular distribution \cite{mitjanpb}. For small values of $N_c$
the matrix elements are correlated and the natural distribution is a
Gaussian \cite{mitjarus}.
On the other hand, from the asymptotic dependence of the matrix
elements of the Dirac operator it was concluded that \cite{Teper}
the spectral density diverges as $1/\sqrt{ \lambda}$. Another estimate
comes from quenched chiral perturbation theory \cite{Golterman}. 
For $\lambda \rightarrow 0$ one obtains a linear dependence 
\cite{Smilga} for $N_f \ne 0$, and a logarithmic divergence \cite{Toublan} for
$N_f =0$. For $N_f \ne 0$ the slope of the Dirac spectral density for 
$\lambda \rightarrow 0$ is consistent with the analytical result 
\cite{acta}. Here we only consider the quenched case.

In order to study the thermodynamic limit of the spectral density we
have unfolded the calculated spectrum with the known result for the 
microscopic spectral density. The resulting 
normalized average spectral density for $N_f = 0$ is shown in Fig. 6. 
Inspection of the normalized eigenvalue density in an interval
$[0.0, 0.05]$ shows that it grows logarithmically with $N$. This suggests
that $\rho(\lambda)$ diverges for $\lambda \rightarrow 0$ in the
thermodynamic limit. In order to understand the nature of this divergence
we have compared results obtained for the 
instanton liquid with the analytical results
for the a chiral version of the $\alpha$-ensenbles \cite{alpha} (dotted curve). 
These are random matrix ensembles with potential
$V(x) = x^{\alpha/2}$ in (\ref{zrandom}). 
The spectral density
of the chiral $\alpha$-ensembles 
diverges logarithmically for $\lambda \rightarrow 0$ at $\alpha =1$. 
The value of $\rho(0)$ is finite for $\alpha > 1$ .
For $N= 64$, 128, 256 and 512 we find $\alpha = 1.217,\, 1.132, \, 1.073,
\, 1.039$ by fitting to the energy interval $[0.05, 0.5]$, respectively. 
In the thermodynamic limit this extrapolates 
to a value of $\alpha = 1.00 \pm 0.02$, resulting in a logarithmic divergence
of the spectral density.
We also compare the average spectral density to a simple power law behavior
$a-b\lambda^\beta$. We find that the spectra are equally well fitted by such
dependence (see full  curve in Fig. 6). 
By fitting to the interval $[0.05, 0.5]$
we find $\beta = 0.28,\, 0.25, \, 0.13, 
\, 0.09$ for $N= 64$, 128, 256, and 512, respectively. This extrapolates
to a value of $\beta =0.03 \pm 0.04$ and is also consistent with a
logarithmic singularity for $\lambda \rightarrow 0$.
We conclude that our results are consistent with an average spectral
density that diverges logarithmically for $\lambda \rightarrow 0$ 
in the limit $V\rightarrow \infty$. However,
larger volumes need to be studied in order to arrive at definite conclusions.
We wish to emphasize that our results 
disagree with estimates obtained by mean field methods.

\begin{center}
\begin{figure}[!ht]
\centering
\includegraphics[width=100mm]{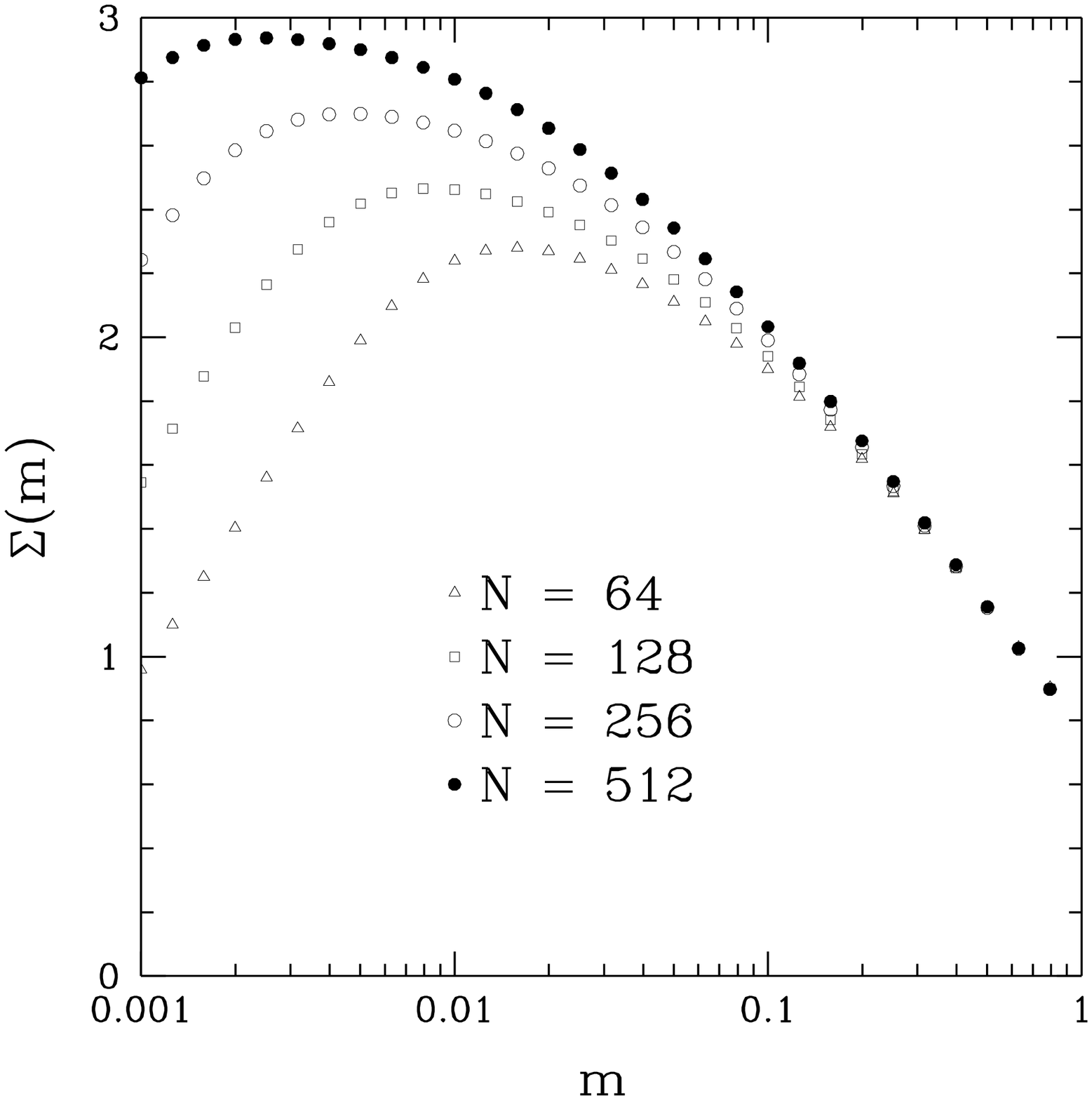}
\vspace*{0.3cm}
\begin{center}
\begin{minipage}{13cm}
\baselineskip=12pt
{\begin{small}
Fig. 7. The valence quark mass dependence of the chiral condensate for
$N_f = 0$ and a total number of instantons as indicated in the legend of
the figure.
\end{small}}
\end{minipage}
\end{center}
\vspace*{-0.1cm}
\end{figure}
\end{center}
The valence quark mass dependence of the chiral condensate defined 
in (\ref{valence})
is shown in Fig. 7. Results are given  for $N_f = 0$ and 
different values of the total number of instantons.
The downward trend for small masses is a finite size effect that can be 
understood analytically in terms of the microscopic spectral density
\cite{vPLB}. The large volume dependence of the "plateau" prevents us from
comparing  our results with 
quenched chiral perturbation theory \cite{Golterman} which predicts a
logarithmic dependence on the valence quark mass. For the same reason,
the connected scalar susceptibility, $\del_m \Sigma(m)$, shows an
even larger volume dependence.

\begin{center}
\begin{figure}[!ht]
\centering
\includegraphics[width=100mm]{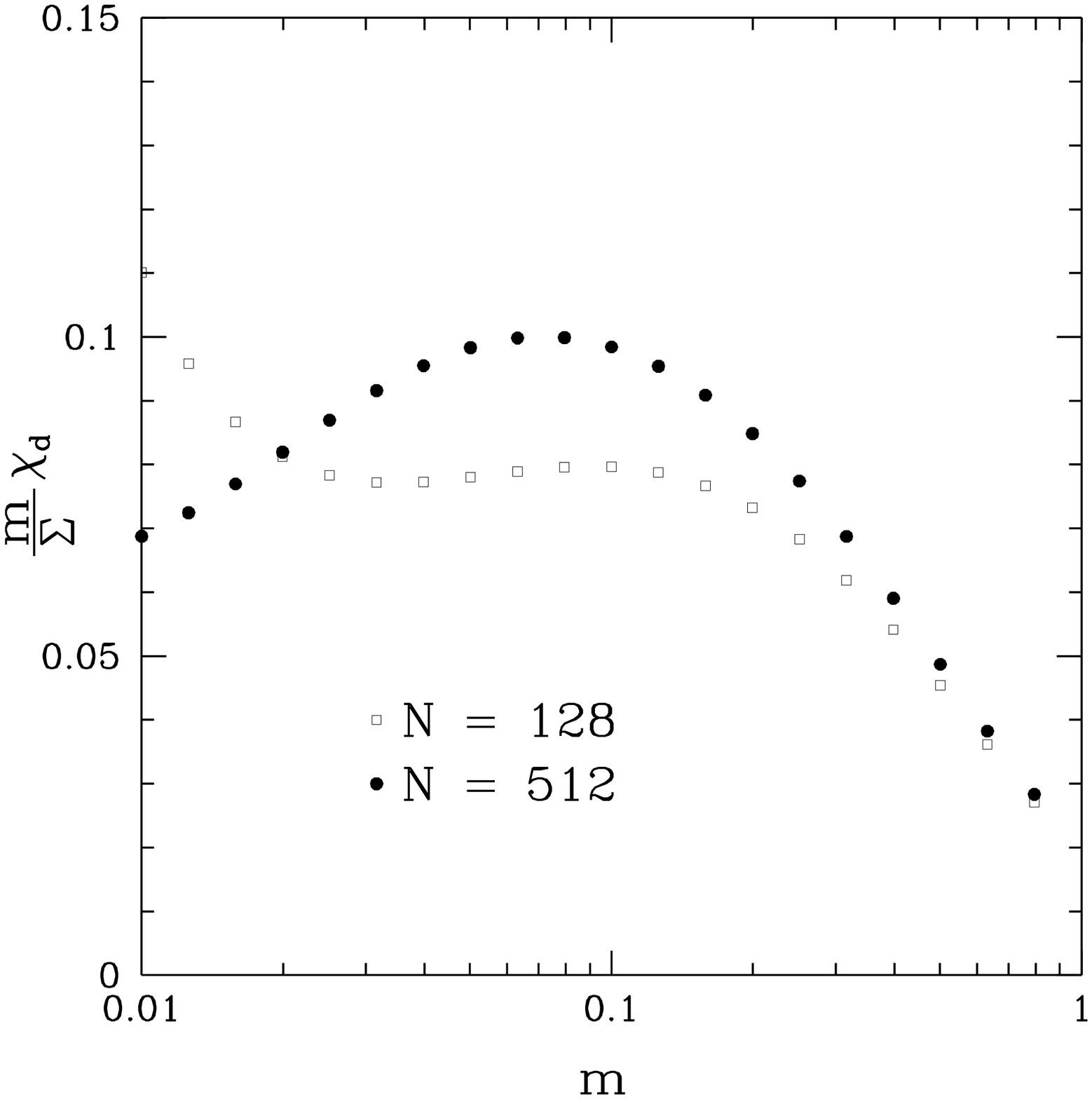}
\vspace*{0.3cm}
\begin{center}
\begin{minipage}{13cm}
\baselineskip=12pt
{\begin{small}
Fig. 8. The disconnected chiral susceptibility for $N_f =0$ both for
128 and 512 instantons.
\end{small}}
\end{minipage}
\end{center}
\vspace*{-0.1cm}
\end{figure}
\end{center}

Finally, we consider the disconnected scalar susceptibility.
It is related to the two-point correlation function according to the formula
\cite{nowak}
\be
\chi_d = \frac{-2}V \int_0^\infty\int_0^\infty d\lambda d\lambda' 
\frac{m_1m_2 (\lambda^2-\lambda'^2)^2 R_c(\lambda,\lambda')}
{(\lambda^2+m_1^2)(\lambda^2+m_2^2)(\lambda'^2+m_1^2)(\lambda'^2+m_2^2)}
\label{chid}
\ee
where $R_c(\lambda,\lambda')$ is the connected two-point correlation
function,
\be
R_c(\lambda,\lambda') = \langle \rho(\lambda) \rho(\lambda') \rangle
-\langle \rho(\lambda) \rangle \langle \rho(\lambda') \rangle.
\ee
In the derivation of (\ref{chid}) we have used that that $R_c$ satisfies the
sum-rule $\int_0^\infty d\lambda R_c(\lambda,\lambda')  = 0$. 
In the bulk of the spectrum, a number variance
given by $\Sigma_2(n) = \chi n$ is reproduced by a two-point correlation
function $R_c(\lambda, \lambda') = \chi (\delta(\lambda-\lambda')\rho(\lambda)
-2\rho(\lambda) \rho(\lambda')/N)$ (For an extensive discussion of the
relation between a linear dependence of the number variance and the sum rule
for the two-point function we refer to \cite{kravtsovmoriond}).
This results in the disconnected susceptibility
\be
\chi_d = \chi \left (\frac{\Sigma}{m} -\del_m \Sigma- 2\Sigma^2\right ).
\ee 
A  $1/m$ behavior for the susceptibility was observed in unquenched 
lattice QCD simulations \cite{Karsch}. 
In Fig. 8 we show the disconnected scalar susceptibility 
$m \chi_d/ \Sigma(m)$ versus $m$. 
We observe a 'plateau' that would indicate a $1/m$ dependence
for $N=128$, but it disappears for larger volumes. Apparently, finite
size effects prevent us from making any definite conclusions.

Let us finally make a naive estimate of the disconnected susceptibility. The
two-point correlation function $T_2(\lambda,\lambda')$, which is related
to $R_c(\lambda,\lambda')$ by $T_2(\lambda,\lambda')= 
\delta(\lambda-\lambda')\rho(\lambda) - R_c(\lambda,\lambda')$,
can be expressed in terms of the two-level cluster
function as 
\be
T_2(\lambda,\lambda') \approx \rho(\lambda)\rho(\lambda') Y_2(r),
\label{T2}
\ee
where $r =\int_\lambda^{\lambda'}\rho(s)ds$. According to \cite{Guhr}
$Y_2(r)$ is related to the number variance as
\be
Y_2(r=n) = -\frac 12 \del_n^2 \Sigma_2(n).
\label{Y2}
\ee
This results in $T_2(\lambda,\lambda') \approx \frac 
12\del_\lambda \del_{\lambda'} \Sigma_2(r)$. 
Combining (\ref{T2}) and (\ref{Y2}) with (\ref{chid}) we find that 
a finite value of $\chi_d$, which is expected according to 
quenched chiral perturbation theory
\cite{Toublan} and lattice QCD simulations \cite{Karsch}, 
requires that
\be
\lim_{N\rightarrow \infty}\frac{ \Sigma_2(Nx)}N 
\ee
is finite. In other words, a nonzero disconnected chiral susceptibility
requires a number variance that shows a linear behavior on macroscopic scales. 
In chiral random matrix theory we find that $\chi_d = 0$ in the
thermodynamic limit.
Because of the large volume dependence of the
spectral density this argument could not be verified in detail. In future
work, though, we hope confirm  relations of this type.

\section{Conclusions}
We have identified an energy scale below which the eigenvalue correlations of
the QCD Dirac operator are given by the chiral random matrix ensembles. In
analogy with the theory of mesoscopic systems, this scale will be called
the Thouless energy. For eigenvalues near zero the volume dependence
of the Thouless energy is weak but its numerical value is of the
same order as estimates from the pion Compton wavelength. 
For eigenvalues in the bulk of
the spectrum the Thouless energy scales roughly with the square root of
the volume which is in agreement with results from the theory of
mesoscopic systems. However, we find a proportionality constant that is
larger than given by our simple estimates. 

For energy scales beyond the Thouless energy 
a linear $n$-dependence of the number variance is found 
which, according to
the work by Altshuler and Shklovskii, corresponds to a critical system.
We have shown that the corresponding Dirac eigenfunctions 
show a multifractal behavior. However,
our multicritical exponents are not in agreement with relations derived
for a critical system. Obviously, our ensemble of instantons is below critical.
One way to increase the criticality of the system is to increase 
the number of flavors. However, at $N_f \ne 0$ 
simulations for the large volumes that are required for the extraction of the
multifractality index became to costly.
Alternatively, one expects that pion loops become important at energy
scales beyond the Thouless energy, and that results from chiral perturbation
theory apply to this region. Due to large finite size effects we 
were unable to make
a detailed comparison. 
However, the spectral density near zero in the quenched 
approximation is consistent with a logarithmic dependence predicted by
quenched chiral perturbation theory. 

\vspace*{0.5cm}
\noindent
{\bf Acknowledgements} We wish to thank A. Smilga for encouraging us to
study localization properties for a liquid of instantons and Y. Fyodorov for
his suggestion that the transition point in our results is related to the
Thouless energy. 
During the completion of this work we received a preprint by 
R. Janik, M. Nowak, G. Papp and I. Zahed
in which similar questions are discussed.

\setlength{\baselineskip}{17pt}

\end{document}